\documentclass[prl,preprint,showpacs,amsmath]{revtex4}
\usepackage{graphicx}
\begin{document}

\title{Universal Critical Behavior of Noisy Coupled Oscillators}

\author{Thomas Risler$^{1,2}$, Jacques Prost$^{3,2}$ and Frank  
J\"ulicher$^{1,2}$}
\affiliation{$^1$Max-Planck-Institut f\"ur Physik komplexer Systeme,
N\"othnitzerstrasse 38, 01187 Dresden, Germany}
\affiliation{$^2$Physicochimie Curie (CNRS-UMR 168), Institut Curie, 26  
rue d'Ulm, 75248 Paris Cedex 05, France}
\affiliation{$^3$Ecole Sup\'erieure de Physique et de Chimie  
Industielles de
la Ville de Paris, 10 rue Vauquelin, 75231 Paris Cedex 05, France}

\date{\today}

\begin{abstract}
We study the universal thermodynamic properties of systems consisting of
many coupled oscillators operating in the vicinity of a homogeneous
oscillating instability. In the
thermodynamic limit,  the Hopf bifurcation
is a dynamic critical point far from equilibrium described by a  
statistical
field theory. We perform a
perturbative renormalization
group study, and show that at the critical point a generic
relation between correlation and response functions appears. At the  
same time the
fluctuation-dissipation relation is strongly violated.
\end{abstract}

\maketitle

The collective behavior of
many interacting elements generally leads to transitions and critical
points in the large-scale and long-time properties of complex systems.  
This is
well understood in the study of systems at thermodynamic
equilibrium \cite{zinnB,brez76c,hohe77}. Non-equilibrium critical  
behaviors
have  been studied in a number of systems
\cite{noneq,schm95,taub02} but remain a serious
challenge.  An important example for criticality far from
thermodynamic equilibrium is the behavior of coupled oscillators in
the vicinity of a continuous homogeneous oscillatory instability or
supercritical Hopf bifurcation.
Such instabilities are important in many physical, chemical and  
biological
systems \cite{stroB,bergB}.

In this paper, we apply concepts of
the theory of dynamic critical points
to study the generic properties of systems of
coupled oscillators in the thermodynamic limit. In particular, we  
discuss linear response and
two-point correlation functions
defined for the oscillator ensemble.
Since the system is far from a thermodynamic equilibrium, the  
fluctuation-dissipation (FD) relation between
correlation and response functions in equilibrium systems
is broken. We  show that a Hopf bifurcation represents a  
non-equilibrium critical point
and study the universal behaviors
characterizing its approach
from the non-oscillating state.
We apply field theoretic renormalization group (RG) methods and develop  
an RG procedure which is appropriate for the case of a spontaneously  
oscillating system.
This RG is performed in an oscillating reference frame with a
scale dependent oscillation frequency. The RG fixed points
characterize the universal critical properties of locally coupled  
oscillators.
We find that at the critical point of a Hopf bifurcation an
FD relation is formally satisfied
if the system is described within the oscillating reference frame. In  
terms of physical variables,
the FD relation is strongly broken but a relationship
between correlation and response functions appears.
Even though our calculations are performed
in a $d=4-\epsilon$ dimensional space, we suggest that the main features
of our results apply to Hopf bifurcations in general.

The generic behavior of an oscillator in the vicinity of a
supercritical
Hopf bifurcation can be described by a dynamic equation for a complex
variable $Z$ which characterizes
the phase and amplitude of the oscillations \cite{stroB}.
This variable can be chosen such that its real part is, to linear order,
related to a physical observable, e.g. the displacement $X(t)$  
generated by a mechanical oscillator:
$X(t)={\rm Re}(Z(t))+$nonlinear terms.
In the presence of a periodic stimulus force $F(t)=\tilde F e^{-i\omega  
t}$
with a frequency $\omega$ close to the oscillation frequency at the  
bifurcation $\omega_0$, the generic dynamics obeys \cite{cama00}
\begin{equation}\label{fx}
\partial_{t}Z=-(r+i\omega_{0})Z-(u+iu_a)\left| Z \right|
^{2}Z +\Lambda^{-1} e^{i\theta} F(t) \quad .
\end{equation}
For $F=0$ and $r>0$,
the static state $Z=0$ is stable. The system undergoes a Hopf  
bifurcation at
$r=0$ and exhibits spontaneous
oscillations for $r<0$.
The nonlinear term characterized by the coefficients
$u$ and $u_a$ stabilizes the oscillation amplitude for $u>0$.
The external stimulus
appears linearly in this
equation and couples in general with a phase shift $\theta$. In
the case of a mechanical oscillator, the coefficient $\Lambda$ has  
units of a friction.
From the point of view of statistical physics, the Hopf bifurcation is a
critical point and Eq. (\ref{fx}) characterizes
the corresponding mean field theory.
Indeed, at $r=0$ and $\omega=\omega_0$ and in terms
of the amplitude $\tilde X$ of the limit cycle $Z(t)=\tilde X  
e^{-i\omega t}$,
the system exhibits a power-law response
$|\tilde X|\simeq | \tilde F|^{1/\delta}$ where $\delta=3$ is a mean  
field critical exponent. For frequency differences
$\vert \omega-\omega_0\vert \gg \Lambda^{-2/3}
\vert \tilde F \vert^{2/3}
\vert u+iu_a \vert^{1/3}$, the response becomes linear with
$\vert \tilde X\vert\simeq \Lambda^{-1} \vert \tilde  
F\vert/\vert\omega-\omega_0\vert$.

In the presence of fluctuations, the critical point of an individual  
oscillator is concealed
in the same way as finite size effects destroy a phase transition in  
equilibrium
thermodynamics. However, a true
critical point can exist in a thermodynamic limit where many  
oscillators,
distributed on a lattice in a $d$-dimensional space,
are coupled by nearest-neighbor interactions.
The combined system undergoes a dynamic phase transition at which all  
oscillators synchronize
and an order parameter, which characterizes the
global phase and amplitude of oscillators, becomes nonzero. Since at  
the critical
point the correlation length diverges and oscillators become  
synchronized over
large distances, a discrete model can on large scales be described by a  
continuum field theory
which characterizes the universal features of the critical point  
\cite{zinnB,fish98}.
In the case of locally coupled oscillators, this field theory is given  
by
the complex Ginzburg-Landau equation \cite{cgleref} with
fluctuations
\begin{eqnarray}\label{cgle}
\partial_{t}Z&=&-(r+i\omega_{0})Z+(c+ic_{a})\Delta Z\nonumber\\
&&-(u+iu_{a})\left| Z \right|^{2}Z+\Lambda^{-1} e^{i\theta} F+\eta  
\quad .
\end{eqnarray}
Here,
$Z(\mathbf{x},t)$ becomes a complex field
defined at positions $\mathbf{x}$ in a $d$-dimensional space
and $\Delta$ denotes the Laplace operator.
The coefficients $c$ and $c_a$ characterize the local coupling of
oscillators and the effects of fluctuations are described via a complex  
random
forcing term $\eta(\mathbf{x},t)$.
For a vanishing external field $F(\mathbf{x},t)$ and in the absence of
fluctuations, Eq. (\ref{cgle}) is invariant with respect to phase  
changes of
the oscillations $Z\rightarrow Ze^{i\phi}$.
As far as long time and
long wavelength properties are concerned, $\eta$ can be chosen
Gaussian and white with the correlations
$\langle\eta(\mathbf{x},t)\eta(\mathbf{x}',t')\rangle=0$ and
$\langle\eta(\mathbf{x},t)\eta^*(\mathbf{x}',t')\rangle=4D\delta^{d}(\mathbf{x}-\mathbf{x}')\delta(t-t')$
which respect phase invariance.

The linear response function $\chi_{\alpha\beta}$ and the two-point  
auto-correlation function
$C_{\alpha\beta}$
of this field theory are defined by
$\langle \psi_\alpha(\mathbf{x},t)\rangle =\int d^d x'\,d  
t'\,\chi_{\alpha\beta}(\mathbf{x}-\mathbf{x}',t-t')
F_\beta(\mathbf{x}',t')+O(\vert F\vert^2)$ and  
$C_{\alpha\beta}(\mathbf{x}-\mathbf{x}',t-t')
=\langle\psi_{\alpha}(\mathbf{x},t)\psi_\beta(\mathbf{x}',t')\rangle_c$.
Here we have expressed $Z=\psi_1+i\psi_2$ and $F=F_1+iF_2$ by their  
real and
imaginary parts and $\langle...\rangle_c$ denotes  a connected
correlation function. Because of phase invariance, these functions obey
symmetry relations, e.g. $C_{11}=C_{22}$, $C_{21}=-C_{12}$.
In the following,we focus for simplicity on the elements
$C\equiv C_{11}$ and $\chi\equiv\chi_{11}$ which characterize the  
correlation
and response of the observable $X$.

It is convenient to
eliminate the frequency $\omega_0$ from Eq. (\ref{cgle}) by
a time-dependent variable transformation $Y\equiv e^{i\omega_0t}Z$,  
$H\equiv
e^{i\omega_0t}\Lambda^{-1} e^{i\theta}F$ and $\zeta\equiv  
e^{i\omega_0t}\eta$.
This leads to the amplitude equation
\begin{equation}\label{cgley}
\partial_{t}Y=-rY+(c+ic_{a})\Delta Y-(u+iu_{a})\left| Y \right|  
^{2}Y+H+\zeta \quad ,
\end{equation}
where
the noise $\zeta$ has the same correlators as $\eta$.
For the particular case $c_a=0$ and $u_a=0$, Eq. (\ref{cgley})
becomes identical to
the model A dynamics of a real Ginzburg-Landau field theory with an  
$O(2)$
symmetry of the order parameter \cite{hohe77}. The critical behavior of  
this
theory at thermodynamic equilibrium has been extensively studied
\cite{modelOn}.
This leads, in this particular case,
to a formal analogy between an equilibrium phase transition and a
Hopf bifurcation.
The correlation and response
functions $C_{\alpha\beta}$ and $\chi_{\alpha\beta}$
can here be obtained from
those of the equilibrium field theory by using the time-dependent  
variable
transformation given above.
Since the theory at thermodynamic equilibrium obeys an FD relation,
a generic relation
between the correlation and response functions $C_{\alpha\beta}$ and
$\chi_{\alpha\beta}$
appears.
This special case provides a further example
of an equilibrium universality class
found in a non-equilibrium
dynamics
with non-conserved order parameter \cite{taub02}.
It is the case e.g.
for the model A dynamics of the real Ginzburg-Landau theory
with a $Z_2$ symmetry \cite{glTheoZ2},
even when the symmetry
is broken by the non-equilibrium perturbations
\cite{bass94}, and for some of its
generalizations
to the $O(n)$ symmetry
\cite{taub97}.

This raises the question
to know
whether the equilibrium universality class
also characterizes the general case where $u_a$ and $c_a$ are finite.
Dimensional analysis reveals that for $d>4$
mean field theory applies. In this case,
\begin{eqnarray}\label{Eq.MFExpr}
\chi^{\rm mf}(\mathbf{q},\omega)&=&\frac{1}{2\Lambda}\left[
\frac{e^{i\theta}}{R-i(\omega-\Omega_0)}+\frac{e^{-i\theta}}{R- 
i(\omega+\Omega_0)}\right]\nonumber\\
C^{\rm
mf}(\mathbf{q},\omega)&=&\frac{D}{R^2+(\omega- 
\Omega_0)^2}+\frac{D}{R^2+(\omega+\Omega_0)^2}
\quad ,
\end{eqnarray}
where $R=r+c\mathbf{q}^2$, $\Omega_{0}=\omega_{0}+c_{a}\mathbf{q}^2$ and
where
$\mathbf{q}$ and $\omega$ are wave vector and angular frequency,
respectively.

For $d<4$ mean field theory breaks down.
We apply RG
methods using an $\epsilon$
expansion  near the upper critical dimension ($d=4-\epsilon$)  
\cite{wils72}.
Defining two real fields $\phi_\alpha$ by
$Y=\phi_{1}+i\phi_{2}$, Eq. (\ref{cgley}) reads
\begin{equation}\label{Eq.MatEvol}
\partial_{t}\phi_{\alpha}=-R_{\alpha\beta}\phi_{\beta}- 
U_{\alpha\beta}\phi_{\beta}\phi_{\gamma}\phi_{\gamma}+H_{\alpha}+\zeta_{ 
\alpha} \quad ,
\end{equation}
where $H=H_1+iH_2$,  
$R_{\alpha\beta}=(r-c\Delta)\delta_{\alpha\beta}- 
c_{a}\Delta\,\varepsilon_{\alpha\beta}$ and $U_{\alpha\beta}=u  
\delta_{\alpha\beta} +
u_a \varepsilon_{\alpha\beta}$, with $\epsilon_{21}=-\epsilon_{12}=1$  
and
$\epsilon_{ij}=0$ for $i=j$.
We introduce the Martin-Siggia-Rose response field
$\tilde{\phi}_{\alpha}$ \cite{mart73} and apply the Janssen-De
Dominicis formalism \cite{Jans_DeDom} to write a generating  functional
with action
\begin{eqnarray}\label{Eq.Action}
\mathcal{S}\left[\tilde{\phi}_{\alpha},\phi_{\alpha}\right]&=&\int d^{d}
x\,d t\, \left\{
D\tilde{\phi}_{\alpha}\tilde{\phi}_{\alpha}- 
\tilde{\phi}_{\alpha}\left[\partial_{t}\phi_{\alpha}+R_{\alpha\beta}\phi 
_{\beta}\right]\right.\nonumber\\
&&\left.\hspace{1cm}-U_{\alpha\beta}\tilde{\phi}_{\alpha}\phi_{\beta}\phi_{\gamma}\phi_{\gamma}\right\} \quad .
\end{eqnarray}

Using a Callan-Symanzik RG scheme \cite{brez76c,zinnB},
we define the renormalized theory such that its effective
action is of the form (\ref{Eq.Action}). This requires to introduce
a phase shift $\delta \theta$  and a frequency shift $\delta\omega_0$
between the bare fields
$(\phi_{\alpha}^0,\tilde{\phi}_{\alpha}^0)$ and the
renormalized fields $(\phi_{\alpha},\tilde{\phi}_{\alpha})$:
\begin{eqnarray}\label{dphidomega}
\phi^0_{\alpha}(\mathbf{x},t^0)&=&\Omega_{\alpha\beta}(-\delta\omega_0  
t)\,Z_{\phi}^{1/2}Z_{\omega}\,\phi_{\beta}(\mathbf{x},t)\nonumber\\
\tilde{\phi}^0_{\alpha}(\mathbf{x},t^0)&=&\Omega_{\alpha\beta}(- 
\delta\theta-\delta\omega_0  
t)\,Z_{\tilde{\phi}}^{1/ 
2}Z_{\omega}\,\tilde{\phi}_{\beta}(\mathbf{x},t) \quad .
\end{eqnarray}
Here we have introduced $Z$-factors for the renormalization of the  
fields and
the time ($t^0=Z_{\omega}^{-1}t$), and
$\Omega_{\alpha\beta}(\theta)$ denotes
the rotation matrix
by an angle $\theta$ in two dimensions.
We furthermore  introduce dimensionless coupling constants $g$ and  
$g_a$ and a scale
factor
$\mu$ by $u=\mu^{\epsilon}(4\pi)^{-\epsilon/2}g$ and
$u_a=\mu^{\epsilon}(4\pi)^{-\epsilon/2}g_a$.
The bare and renormalized quantities are now related depending on $\mu$
\footnote{Note that $c$ and $D$ are not
renormalized and that we choose units such that  $c=1$ in the  
following.}.
We define  the correlation and
response functions  
$G_{\alpha\beta}=\langle\phi_\alpha\phi_\beta\rangle_c$ and  
$\gamma_{\alpha\beta}=\langle\phi_\alpha\tilde{\phi}_\beta\rangle_c$.
They are related to the physical observables $C_{\alpha\beta}$ and  
$\chi_{\alpha\beta}$
via \footnote{Note that the frequency $\omega_0$ and the phase $\theta$  
in
Eq.  (\ref{Eq.Phys/MathCoRespFct}) are renormalized and satisfy
$\omega_0=Z_{\omega}^{-1}\omega_0^0+\delta\omega_0$ and  
$\theta=\theta^0+\delta\theta$.}
\begin{eqnarray}\label{Eq.Phys/MathCoRespFct}
C_{\alpha\beta}(\mathbf{x},t^0)&=&\Omega_{\alpha\sigma}\left(- 
\omega_0\,t\right)Z_{\phi}Z_{\omega}^{2}\,G_{\sigma\beta}(\mathbf{x},t)\nonumber\\
\chi_{\alpha\beta}(\mathbf{x},t^0)&=&\Lambda^{ 
-1}\,\Omega_{\alpha\sigma}\left(\theta- 
\omega_0\,t\right)(Z_{\phi}Z_{\tilde{\phi}})^{1/ 
2}Z_{\omega}^{2}\,\gamma_{\sigma\beta}(\mathbf{x}, t) \quad .
\end{eqnarray}
The dependence of the renormalized  parameters $g$, $g_a$ and $c_a$ on  
$\mu$
defines three $\beta$-functions. Writing  
$\vec{g}=\left(g,g_a,c_a\right)$, we
have
$\vec{\beta}\left(\vec{g},\epsilon\right)=\mu(\partial_{\mu}  
\vec{g})_0$, where
$\vec{\beta}=\left(\beta,\beta_a,\beta_{c}\right)$ and  
$(\partial_{\mu})_0$
denotes differentiation with fixed $u^0$, $u^0_a$ and $c^0_a$.

\begin{figure}[h]
\scalebox{0.8}{
\includegraphics{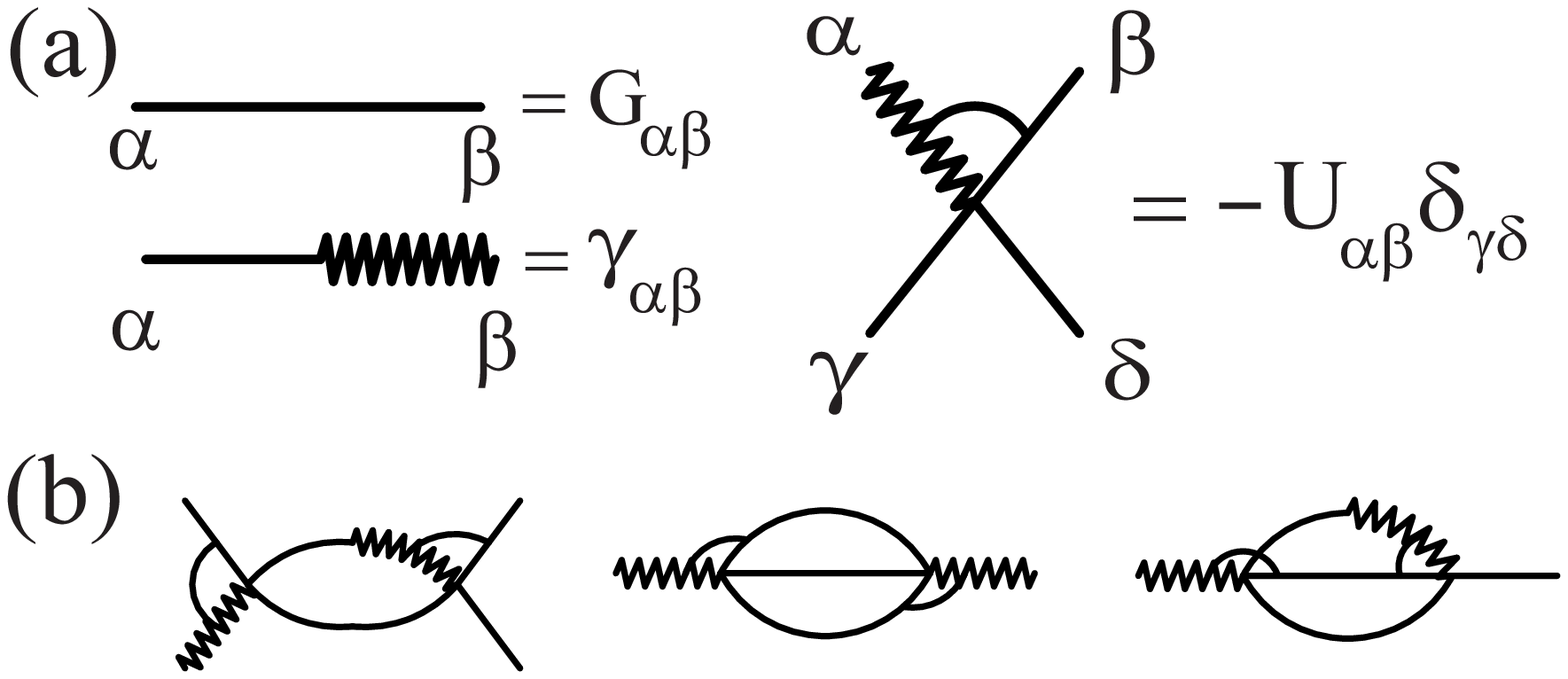}
}
\caption[FeynDiag]
{\label{Fig.FeynDiag}
(a) Graphic representation of the propagators $G_{\alpha\beta}$ and
$\gamma_{\alpha\beta}$ and of the vertex
$U_{\alpha\beta}\delta_{\gamma\delta}$. (b) Examples of Feynman diagrams of
the theory to one and two loop order.}
\end{figure}

To one-loop order in
perturbation theory (see Fig. \ref{Fig.FeynDiag} for examples of Feynman
diagrams of the theory), only
$\omega_0$, $r$, $g$ and $g_a$ are renormalized.
The two non-trivial $\beta$-functions are given by:
\begin{eqnarray}\label{Eq.BandBaFctCS}
\beta
&\simeq&-\epsilon g-\mathcal{D}\left[\frac{g_a^2-g^2-2gg_a  
{c}_a}{1+{c}_a^2}-4g^2\right]\nonumber\\
\beta_a
&\simeq&-\epsilon  
g_a-\mathcal{D}\left[\frac{{c}_a}{1+{c}_a^2}(g^2-g_a^2+2gg_a
{c}_a)-6gg_a\right] \quad ,
\end{eqnarray}
where $\mathcal{D}=4D/(4\pi)^2$.
The RG fixed point corresponds to the values $\vec{g}^*$
of the parameters $\vec{g}$ for which the three
$\beta$-functions are simultaneously zero.
Since $\beta_c=0$ to one-loop order, one condition is lacking
to fully determine
$\vec{g}^*$.
Choosing $c_a$ as a parameter, we obtain $g^*\simeq  
\,\epsilon/5\mathcal{D}$
and $g_a^*\simeq \,c_a\,\epsilon/5\mathcal{D}$.
In order to determine completely the fixed point $\vec{g}^*$,
we need to go to two-loop
order in perturbation theory. To this order all parameters (apart from  
$c$ and
$D$), the fields and the time are renormalized and explicit expressions  
for
all the Wilson's functions and $Z$-factors of the theory can be obtained.

Only one fixed point exists which
describes the universality class of  Hopf bifurcations. It obeys
$c_a^*=0$, $g_a^*=0$ and is infrared-stable. This fixed point
is formally equivalent to the one of
the real Ginzburg-Landau theory with $O(2)$ symmetry.
As a consequence, we find $\nu\simeq1/2+\epsilon/10$,  
$\eta\simeq\epsilon^2/50$ and
$z\simeq2+\epsilon^2\left(6\ln(4/3)-1\right)/50$ which are
the corresponding equilibrium critical exponents.
Here, $\nu$ denotes the exponent characterizing
the divergence of the correlation length $\xi$, $z$ is the dynamic exponent  
and $\eta$ denotes the exponent characterizing the field
renormalization \cite{hohe77,zinnB}.

The RG flow in the vicinity of the fixed point however is defined here  
in a larger
parameter space as the one
corresponding to
the $O(2)$ dynamic model. Furthermore, the effective
theory discussed here is expressed in an
oscillating reference frame  with scale-dependent frequency and phase.
Therefore the correlation and response functions $C_{\alpha\beta}$ and  
$\chi_{\alpha\beta}$
differ from those of
the equilibrium model
and additional universal exponents appear.
We can derive generic expressions for these functions
using the RG flow of all parameters in the vicinity of the
critical point and employing a matching procedure \cite{rudn76}.
For example, we find for $q\xi\gg 1$ and for stimulation at the  
effective oscillation frequency
$\omega_0^{\rm eff}$:
\begin{equation}
\chi(q,\omega=\omega_0^{\rm{eff}})\simeq\frac{1}{q^{2- 
\eta}}\,\frac{1}{2\Lambda_{\rm{eff}}}\left[\frac{e^{i\theta(q)}}{1+i\gamma(q)}\right] \quad ,
\end{equation}
where we have introduced the functions
$\theta(q)\simeq \theta_{\rm{eff}}+
\alpha_{\rm{eff}}q^{\omega_1}+\beta_{\rm{eff}}q^{\omega_2}$ and
$\gamma(q)\simeq \gamma_{\rm{eff}}q^{\omega_2}$, and non-universal  
effective
quantities denoted by the index ``eff''. The universal exponents
$\omega_1\simeq \epsilon/5$ and $\omega_2\simeq \epsilon^2/50$
here are  characteristic for a Hopf bifurcation.
Similarly, we find expressions for the correlation function:
\begin{equation}
C(q,\omega=\omega_0^{\rm{eff}})\simeq\frac{1}{q^{z+2- 
\eta}}\,\frac{D_{\rm{eff}}}{1+\gamma(q)^2} \quad ,
\end{equation}
and for the frequency dependence of the homogeneous mode $q=0$
in the regime $(\omega-\omega_0^{\rm{eff}})\xi^z\gg  
1$\cite{risltobepublished}.

Because of the formal analogy of the RG fixed point discussed here with
the one of an equilibrium field theory,
an FD relation appears exactly at the critical point
and relates the functions $G_{\alpha\beta}$ and $\gamma_{\alpha\beta}$.
Since the physical correlation and response functions
$C_{\alpha\beta}$ and $\chi_{\alpha\beta}$ can be determined from
$G_{\alpha\beta}$ and $\gamma_{\alpha\beta}$ using the scale dependent
variable transformations
of Eq. (\ref{Eq.Phys/MathCoRespFct}), a
relation between correlation and response functions appears:
\begin{eqnarray}
\cos{\theta_{\rm{eff}}}\chi''_{11}- 
\sin{\theta_{\rm{eff}}}\chi''_{12}&=&\frac{1}{2\Lambda_{\rm{eff}}D_{\rm{ 
eff}}}\left(\omega
C_{11}+i\omega_0^{\rm{eff}}C_{12}\right)\nonumber\\
\cos{\theta_{\rm{eff}}}\chi'_{12}+\sin{\theta_{\rm{eff}}}\chi'_{11}&=
&\frac{1}{2\Lambda_{\rm{eff}}D_{\rm{eff}}}\left(\omega_0^{\rm{eff}}
C_{11}+i\omega C_{12}\right) \quad . \label{genrel}
\end{eqnarray}
Here, $\chi_{\alpha\beta}=\chi_{\alpha\beta}'+i\chi_{\alpha\beta}''$  
has been
separated in its real and imaginary parts.
The relation (\ref{genrel}) is asymptotically satisfied in the long  
time and wave-length
limits at the critical point.

The physical correlation and response functions $C_{\alpha\beta}$ and  
$\chi_{\alpha\beta}$
however do not obey the equilibrium FD relation. The degree of this  
violation
can be characterized by a frequency-dependent
effective temperature $T_{\rm eff}$:
\begin{equation}
\frac{T_{\rm eff}(\omega)}{T}=\frac{\omega}{2 k_B  
T}\,\frac{C_{11}(\omega,q=0)}{\chi''_{11}(\omega,q=0)} \quad .
\end{equation}
Here, $k_B$ denotes the Boltzmann constant and $T$ is the temperature
of the system.
We find that $T_{\rm eff}/T\sim (\omega-\omega_0^{\rm eff})^{-\sigma}$
diverges at the critical point with a universal exponent $\sigma$. For  
the
particular case $c_a=0$ and $u_a=0$,  $\sigma=1$, while otherwise
$\sigma\simeq 1-\epsilon/5$ to first order in $\epsilon$.
The power-law divergence of $T_{\rm eff}$ as a function of frequency  
reveals a violent
breaking of the FD relation.
This divergence at the oscillation frequency
has been experimentally observed on a single
active oscillating system \cite{mart01b}.

We have shown that the critical point in $d=4-\epsilon$ dimensions of
locally coupled oscillators is formally related to the equilibrium
phase transition in the $XY$ model.
 From this analogy it follows that on the oscillating side
of the Hopf bifurcation, the system of coupled oscillators exhibits  
long range phase order and coherent oscillations for $d>2$.
We can speculate how our results are modified in lower dimensions $d$.
In analogy with the equilibrium $XY$ model, we expect
the phase order of the oscillations
to vanish for $d<2$, and to be quasi-long range exactly at the lower  
critical dimension $d=2$.
In the last case, spectral peaks
on the oscillating side of the Hopf bifurcation
are expected to exhibit power-law tails
with non-universal exponents. If the formal analogy with the  
equilibrium critical point
found here in $d=4-\epsilon$ persists in $d=2$, we would expect to see  
features of
the Kosterlitz-Thouless universality class \cite{kost73} in systems
of coupled oscillators in this dimension.

Critical oscillators are ideally suited for nonlinear signal detection  
and amplification.
Indeed, close to the critical frequency, the linear
response function exhibits divergent behaviors,
indicative of a high sensitivity of the system.
It has been suggested that the ear of vertebrates uses critical  
oscillations
of mechanosensitive hair cells
for the detection of weak sounds and that the properties of the  
critical point
provide the basis
to explain
the observed compressive nonlinear response to mechanical stimuli
and to frequency selectivity in the ear \cite{cama00,egui00}.  The  
correlation and
response functions of single mechanosensory hair bundles have been  
determined
experimentally \cite{mart01b}. These single cell experiments detected  
vibrations at
the scale of tens of nanometers.
There, the Hopf bifurcation was concealed by
finite-size effects
but its signature could be observed.
In the cochlea of mammals, power-law responses over several orders of  
magnitude have been
seen \cite{rugg97}.
This suggests that in such systems a large number of
oscillating degrees of freedom operate collectively and bring the  
system closer to true criticality.

The critical oscillations discussed here can in principle
be realized in artificial systems. Nanotechnology aims to build  
functional units
on the sub-micrometer scale. Large arrays of nano-rotators or  
oscillators on
patterned substrates coupled to their neighbors by elastic or viscous  
effects
would provide a 2-dimensional realization of our field theory.  This  
could permit
in the future experimental studies of the critical phenomena discussed  
here.

We thank Edouard Br\'ezin, Erwin Frey and Kay Wiese for useful  
discussions.


\end{document}